\documentclass{article}
\usepackage{arxiv}

\usepackage{CJKutf8}
\usepackage[utf8]{inputenc} 
\usepackage[T1]{fontenc}    
\usepackage{hyperref}       
\usepackage{url}            
\usepackage{booktabs}       
\usepackage{amsfonts}       
\usepackage{nicefrac}       
\usepackage{microtype}      
\usepackage{lipsum}
\usepackage{graphicx}
\graphicspath{ {./images/} }

\usepackage{listings}

\begin{document}

\title{\LARGE Tracing patient PLOD by mobile phones\\
\large Mitigation of epidemic risks based on patient locational open data}

\author{
 Ikki Ohmukai \\
  Graduate School of Humanities and Sociology\\
  The University of Tokyo, Tokyo, Japan \\
  \texttt{i2k@l.u-tokyo.ac.jp}\\
   \And
 Yasunori Yamamoto\\
  Database Center for Life Science\\
  Research Organization of Information and Systems, Tokyo, Japan \\
  \texttt{yy@dbcls.rois.ac.jp}\\
  \And
 Maori Ito \\
  PLOD info, Japan\\
  \texttt{maorinphone@plod.info} \\
  \AND
 Takashi Okumura\\
  National Institute of Public Health, Japan\\
  Kitami Institute of Technology, Kitami, Hokkaido, Japan\\
  \texttt{tokumura@mail.kitami-it.ac.jp}
}

\maketitle

\begin{abstract}

In the cases when public health authorities confirm a patient with highly contagious disease, they release the summaries about patient locations and travel information.
However, due to privacy concerns, these releases do not include the detailed data and typically comprise the information only about commercial facilities and public transportation used by the patients.
We addressed this problem and proposed to release the patient location data as open data represented in a structured form of the information described in press releases.
Therefore, residents would be able to use these data for automated estimation of the potential risks of contacts combined with the location information stored in their mobile phones.
This paper proposes the design of the open data based on Resource Description Framework (RDF), and performs a preliminary evaluation of the first draft of the specification followed by a discussion on possible future directions.\\
\end{abstract}

\section{Introduction}

\begin{figure*}
\centering
\includegraphics[width=14.0cm]{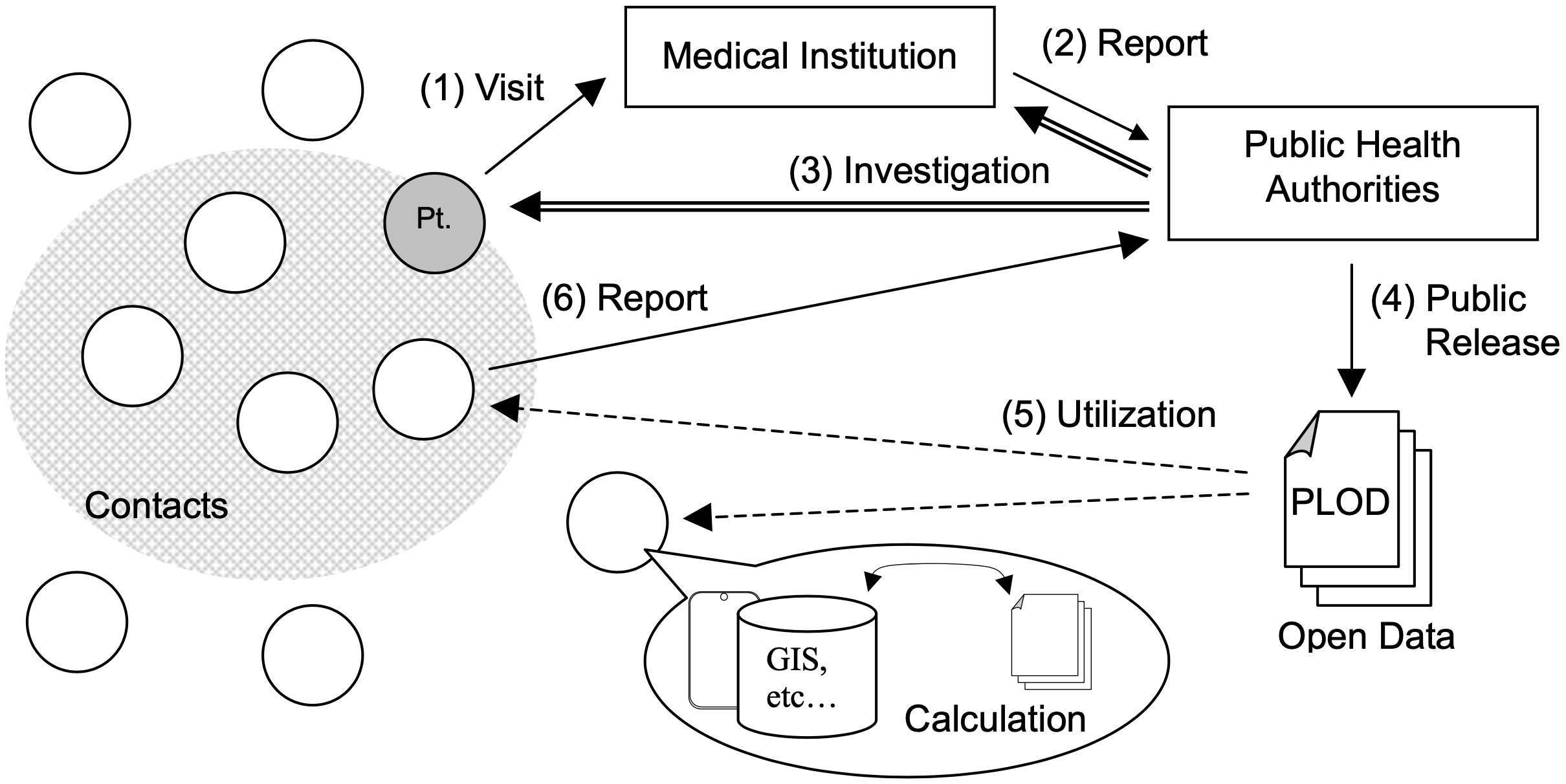}
\caption{Overview of the proposed approach}
  \vspace*{-4mm}
\label{fig:overview}
\end{figure*}

In the cases when public health authorities confirm a patient with highly contagious disease, such as measles, tuberculosis, and the novel coronavirus, they have to identify close contacts of infected patients to monitor their health conditions for a certain period \cite{rasmussen2018cdc}.

This {\it contact tracing} is mostly performed manually by officials.
However, concerning the usage of public transportation and visiting public facilities, the officials are unable to trace all possible contacts.
To mitigate the potential risks associated with such situations, public health authorities release the reports on the travel information of the infected patients, expecting potential contacts to report themselves.
However, due to privacy concerns, authorities cannot provide the detailed information about the patients, and therefore, these releases typically comprise the data related to commercial facilities and public transportation used by the patients.

Most residents rarely pay attention to such releases, except the families who have members in immunocompromised conditions, for example.
Newspapers and TV programs might broadcast the releases; however, such news usually have a limited number of readers or audiences, and therefore, reach only a small part of the population.

Accordingly, public health authorities are not capable of efficiently alerting the potential contacts about the risk associated with the infectious agents, who are mostly unconcerned.
We consider that sharing the open data on the patient location information can be used to efficiently bridge this gap by {\em personalizing the alerts for each resident through the automated estimation of potential intersections between the resident and the infected patients} based on the location history stored in their own mobile phones.
Such approach can be helpful, particularly, when patients have visited various places, and when the number of patients has increases to a certain threshold exceeding the capacity of ordinary people to process the trace information corresponding to all patients.

The proposed open data framework for tracing the patient information is illustrated in Figure~\ref{fig:overview}.
First, a patient visits a medical institution and a physician confirms the infection case (1).
Secondly, the physician reports to a regional public health center (2).
Third, the officials perform the necessary investigation (3) and publicly alert the information by an announcement (4).
In the proposed scheme, this announcement is released as the open data established in a form of
the Patient Locational Open Data (PLOD).
These data can be publicly used to perform mapping of patient traces and estimate the risks of potential contacts for individuals (5).
To perform this estimation, residents can employ the location history information on their mobile phones.
A suspicious contact is supposed to be reported to the public health authority with the purpose of obtaining further instructions (6).

The rest of the paper is organized as follows.
First, Section~\ref{sec:related} provides an overview on the related research focused on the use of the mobile location information
in the public health field.
Section~\ref{sec:design} outlines the design of the proposed open data scheme.
Section~\ref{sec:analysis} demonstrates the preliminary evaluation of the proposed approach, followed by Section~\ref{sec:discussion} to discuss the advantages and limitations.
Section~\ref{sec:conclusion} concludes the paper.

\section{Related works}
\label{sec:related}

Utilization of the mobile location information for the public purposes was initiated in the late 2000s, when mobile phones became commonly available across the world \cite{yang2009use}.
It was deemed as a straightforward approach to trace patient by analyzing the location information provided by mobile devices.
Such application was first realized as a countermeasure for the epidemic of Middle East Respiratory Syndrome (MERS) in 2015, reportedly \cite{kim2017middle}.
However, the trace information is considered as the privacy data, which is highly sensitive concerning individuals, and therefore, there are ethical concerns even considering public health applications \cite{jones2019toward}.
Due to the privacy issue, it has been hardly possible to utilize the personal information to facilitate public health preparedness, and only a limited number of related research works have been published.

At the same time, the research works on the statistical utilization of the mobile location information have gained wide popularity, as mobile phones produced a vast amount of the location information relevant for the improvement of population health \cite{ouedraogo2019does,lai2019measuring,jones2018challenges,jahani2017improving}.
In this regard, the statistical utilization of the mobile location information to control infectious diseases has been considered as a promising approach \cite{chirombo2018review,panigutti2017assessing,sallah2017mathematical,wesolowski2016connecting}.
This field of research has been investigated when the location data related to phone calls were used for efficient surveillance of post-disaster situations and for managing the quarantine of cholera \cite{bengtsson2011improved,finger2016mobile,cinnamon2016evidence}, which was then applied to modeling the spread of infectious diseases \cite{bengtsson2015using}.

It should be noted that these studies utilized the caller-receiver information of mobile devices referred to as Call Detail Records (CDR).
In the related studies, researchers basically used the anonymized or statistically processed data.
Nevertheless, there were concerns about privacy issues even in these cases, and the related studies were mostly conducted in developing countries where the associated regulations were less strict.
In such countries, mobile phones became common owing to their efficiency as social infrastructure means, and healthcare applications based on such devices also became popular, for example, such as m-health \cite{gagnon2016m}.
However, in developing countries, the majority of population use prepaid devices, and therefore, mobile carriers do not possess subscriber contracts.
Accordingly, the trace data are not accompanied with the demographic information, which is a severe limitation associated with this approach \cite{wesolowski2016connecting}.

In 2020, the situation has drastically changed due to the pandemic of the new coronavirus infectious disease (COVID-19).
In China, the government authority have deployed the facial recognition technology, security cameras, and social media analysis combined with the manual surveillance to automate the contact tracing with regard to COVID-19 patients \cite{WSJfeb52020}.
For this purpose, the government have collected necessary records of calls and the location information provided by transportation and communication companies, which are owned by the government.
In South Korea, public health authorities have monitored the credit card history, security camera footage, location information obtained from mobile devices, public transportation cards, immigration records, etc., to track individuals exposed to the risks of possible infection \cite{WSJfeb172020}.
In Taiwan, the subjects of the home quarantine have been monitored by the authority, reportedly, through analyzing mobile phone signals \cite{WSJfeb172020}.
Such applications of processing the location information with the purpose of crisis management of infectious diseases have been rarely published as research articles before the novel coronavirus pandemic in 2020, except for the cases of patient tracing in South Korea \cite{kim2017middle} and a proposal on a risk assessment service of potential infections \cite{okumura2019ghtc}.

The challenge associated with the knowledge representation of the information about infected patients and their location data lies in the fact that such approach should represent both the qualitative information corresponding to attributes and conditions of patients and quantitative information on the location, such as coordinates, distance, and range.
Conventionally, these data need to be handled by authorized independent systems.
In this regard, analyzing the geospatial information has gained less attention in the information retrieval field.
Even concerning the systems with an extension to perform geospatial estimations, the interoperability between systems is not guaranteed \cite{torniai2006sharing}.
To address the problem, the World Wide Web Consortium (W3C) proposed a framework that can consistently handle qualitative and quantitative information in in terms of geospatial knowledge called GeoSPARQL \cite{battle2011geosparql}.
GeoSPARQL has been implemented as an extension to the resource description framework (RDF) \cite{beckett2004rdf} and SPARQL \cite{prudhommeaux2008sparql}, a query language available in RDF, which was standardized by the W3C.
GeoSPARQL can be used to realize the representation and retrieval of information in terms of both qualitative and quantitative data and to enable inference operations on the knowledge, providing the foundation for facilitating the representation of patient tracing.

\section{Design}
\label{sec:design}

In this section, we propose a structured representation of the press releases appropriate to publish the information as PLOD.
The press releases representing the current situation about infected patients usually comprise the following information formulated in natural language, which is not suitable for automated processing.

\begin{itemize}
    \item Issuer and contact information of the document;
    \item Patient's attributes (age and gender);
    \item Name of the disease (infected agents) and infectivity;
    \item Other medical background;
    \item Clinical course of the patient;
    \item Travel information.
\end{itemize}

The metadata include the name of the government agency or local government,
as a publisher of the releases, date and time of the publication.
The patient information includes demographic attributes such as age, gender, occupation, and place of residence, as well as the information on the disease, such as the infectious agents, and the degree of infectivity.
Other medical background and clinical course include the date of onset, confirmed date, and past medical history.
Outline of the travel history might be included here as well.
The releases contain the information to provide an overview on the patient situation; however, the amount of the information included in a particular release may vary. It may describe a single case or represent only the number of patients infected by a certain disease.

\begin{figure*}
\centering
\vspace*{5mm}
\includegraphics[width=16.0cm]{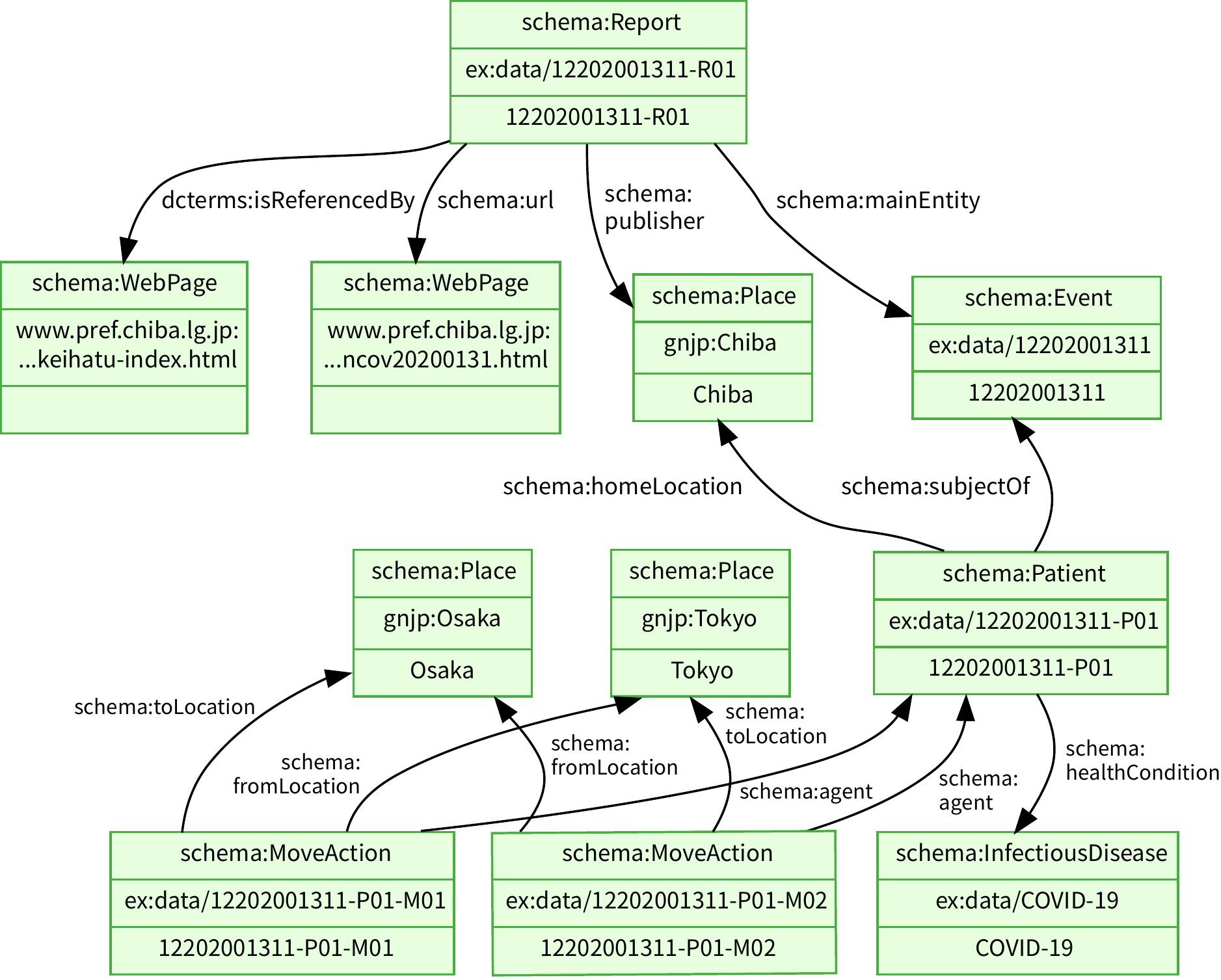}
\vspace*{3mm}
\caption{RDF model of PLOD}
\label{fig:rdfmodel}
\vspace*{0mm}
\end{figure*}

The travel information of patients is mostly composed of a combination of the following items:
i) Travel time or range;
ii-a) Name of the facility, or address visited by the patient;
or ii-b) The list of public transportation used.
In some cases, it may come in a negation form due to denying a visit of a specific place. Figure~\ref{fig:rdfmodel} represents an example of a press release that outlines the two travels of a patient: from Tokyo to Osaka, and from Osaka to Tokyo, as described in the PLOD model.

To model the information, unique and persistent identifiers (PIDs) are required.
In the proposed model, each entity has a persistent ID.
An event entity is automatically generated based on the press release ID so that all elements are connected through the event entity, even when it is not explicitly mentioned in the press release.
Patient IDs are constructed based on the event ID adding a unique number of a the patient.
To identify the travel information, their IDs are generated by adding an incremental number in the travel history associated with a patient ID.
The PLOD URIs are determined uniformly from these IDs by using the HTTP scheme (https://), domain name (plod.info), and path information (/data/) as a prefix.

Most of the properties corresponding to each entity employ Schema.org.
Missing attributes in Schema.org are originally defined with the prefix "ex:".
We select the following properties to represent the patient travel information that is the most important part of the PLOD: "schema:startTime", "schema:endTime", "schema:fromLocation", and "schema:toLocation" defined in the class "schema:MoveAction".
The authenticity is ensured by specifying a URI of the original file issued by the local public health authority.
Figure \ref{fig:turtle} represents the raw data of the PLOD model using the turtle format.

\begin{CJK}{UTF8}{ipxm}
\begin{figure*}
\centering
\begin{lstlisting}
@prefix rdf: <http://www.w3.org/1999/02/22-rdf-syntax-ns#> .
@prefix rdfs: <http://www.w3.org/2000/01/rdf-schema#> .
@prefix schema: <https://schema.org/> .
@prefix dcterms: <http://purl.org/dc/terms/> .
@prefix foaf: <http://xmlns.com/foaf/0.1/> .
@prefix gnjp: <http://geonames.jp/resource/> .
@prefix plod: <https://plod.info/property/> .

<https://plod.info/data/12202001311> a schema:Event ;
    rdfs:label "12202001311" .

<https://plod.info/data/12202001311-R01> a schema:Report ;
    rdfs:label "12202001311-R01" ;
    schema:mainEntity <https://plod.info/data/12202001311> ;
    plod:numberOfPatients "1"^^schema:Integer ;
    schema:datePublished "2020-01-31"^^schema:DateTime ;
    schema:publisher gnjp:Chiba ;
    schema:url <https://www.pref.chiba.lg.jp/shippei/press/2019/ncov20200131.html>;
    dcterms:isReferencedBy <https://www.pref.chiba.lg.jp/shippei/kansenshou/keihatu-index.html>.

<https://plod.info/data/12202001311-P01> a schema:Patient ;
    rdfs:label "12202001311-P01" ;
    schema:subjectOf <https://plod.info/data/12202001311> ;
    schema:healthCondition <https://plod.info/entity/COVID-19> ;
    plod:dateConfirmed "2020-01-31"^^schema:DateTime ;
    foaf:age "20s" ;
    schema:gender "Female" ;
    schema:homeLocation gnjp:Chiba .

<https://plod.info/data/12202001311-P01-M01> a schema:MoveAction ;
    rdfs:label "12202001311-P01-M01" ;
    schema:agent <https://plod.info/data/12202001311-P01> ;
    schema:startTime "2020-01-16"^^schema:DateTime ;
    schema:endTime "2020-01-16"^^schema:DateTime ;
    schema:fromLocation gnjp:Tokyo ;
    schema:toLocation gnjp:Osaka ;
    schema:instrument "Airplane"@ja .

<https://plod.info/data/12202001311-P01-M02> a schema:MoveAction ;
    rdfs:label "12202001311-P01-M02" ;
    schema:agent <https://plod.info/data/12202001311-P01> ;
    schema:startTime "2020-01-22"^^schema:DateTime ;
    schema:endTime "2020-01-22"^^schema:DateTime ;
    schema:fromLocation gnjp:Osaka ;
    schema:toLocation gnjp:Tokyo ;
    schema:instrument "Bus"@ja .

<http://geonames.jp/resource/Tokyo> a schema:Place ;
    rdfs:label "Tokyo" .

<http://geonames.jp/resource/Osaka> a schema:Place ;
    rdfs:label "Osaka" .

<http://geonames.jp/resource/Chiba> a schema:Place ;
    rdfs:label "Chiba" .
    
<https://plod.info/entity/COVID-19> a schema:InfectiousDisease ;
    rdfs:label "COVID-19" ;
    schema:name "2019-nCoV acute respiratory disease"@en ;
    schema:infectiousAgent "2019-nCoV" ;
    schema:code <http://purl.bioontology.org/ontology/ICD10/U07.1> .

<http://purl.bioontology.org/ontology/ICD10/U07.1> a schema:MedicalCode ;
        schema:codeValue "U07.1" ;
        schema:codingSystem "ICD-10" .

\end{lstlisting}
\caption{Example of PLOD (excerpt)}
\label{fig:turtle}
\vspace*{-3mm}
\end{figure*}
\end{CJK}

\section{Analysis}
\label{sec:analysis}

\subsection{Analysis of press releases for infectious diseases}

To verify the appropriateness of the patient information described in the PLOD model, we collected the press releases corresponding to infectious disease patients published by the local governments in Japan.
The survey results included the press releases of 189 measles cases and 39 COVID-19 ones starting from 2018 through 2020, in total, 228 cases published by the Ministry of Health, Labor, and Welfare, as well as prefectures and cities.
We then categorized the documents according to the way used to describe the case trace information.
The obtained results are summarized in Table~\ref{tab:categories}.

Among the 228 releases, 71 described the trace in the independent sections, while 9 releases mentioned the trace in the bodies of the announcements.
Another 80 releases included the trace information using generic terms and avoided providing the specific location.
Then, 16 releases notified that there were no risks of public exposure by the cases, and 45 releases did not provide any detailed information of the case.
In sum, among of 228 releases, 224 contained the case descriptions and 80 included the trace information.
The format of the press releases and the contents were not uniform: some of them included the names of specific places, and others did not present the information sufficiently detailed to trace the patients.

\begin{table}
\vspace*{6mm}
\caption{Press-release of the infected cases published by local governments in Japan}
\vspace*{5mm}
\centering
  \begin{tabular}{lrrrr}
&   2018&   2019&   2020&   Total\\ \hline
Detailed description of trace&   3&   67&  1&  71\\ 
Mentions of trace&  1&  3&  5&  9\\ 
Non-specific description of trace&  30&  20&  30&  80\\ 
No public exposure&  1&  15&  0&  16\\ 
Not available&  16&  26&  3&  45\\ 
Others&  0&  3&  0&  3\\
No case description&  3&  1&  0&  4\\ \hline
Yearly total&  54&  135&  39&  228\\
  \end{tabular}
\label{tab:categories}
\vspace*{4mm}
\end{table}

Names of the hospitals visited by the patients were in some cases omitted in the announcements.
Medical institutions kept the patient records as per each day, and public health authorities and administrators of the institutions could directly contact the possible contacts.
Accordingly, in Japan, such information could have been excluded from the releases intentionally, for privacy reasons.

The survey is incomplete in the sense that the analysis has been conducted only based on the releases found by using search engines, and there are releases that can be found only in archives of the local governments.
Nevertheless, it is highly probable that the press releases corresponding to the patients infected by contagious agents do not have a single standardized format.
In this regard, announcements by public health authorities in English-speaking countries are deemed to have similar characteristics, concluding from the twelve announcements we found by using search engines.
The tentative conclusion can be summarized as follows: the information provided by public health authorities is not standardized, and it is considered meaningful to propose a framework for the unified representation of the actual information about infected patients.

\subsection{Evaluation of the proposed model}

Next, we evaluate the expressive power of the proposed PLOD model based on the press releases that contain the detailed trace information about patients.
For the evaluation, we analyzed 39 press releases issued in the period from Jan 16, 2020 to Feb 11, 2020 and arranged the available information according to the
PLOD model by using RDF (in the turtle format).
The metadata of the press release itself, the patient information, and the patient trace information were successfully expressed for all cases by using PIDs issued for each entity considered as a Subject.
However, we confirmed that the information corresponding to Objects became empty for the cases in which the detailed data were not provided in the corresponding release.
Among the considered 39 press releases, 23 included the patient information, and 8 out of those 23 represented the trace information of patients, including date and location.

To estimate the contact risk using the PLOD model, it is essential to include the origin of a trace or a destination, preferably, together with the detailed information about the visited places.
To obtain the detailed description of the trace information, we extracted the steps from each case report, and 21 steps were identified on the basis of the 8 releases.
Among these steps, 10 included the origin or destination data with varying granularity: 8 were described at the prefecture level, 1 as ``oversea'', and only 1 contained the facility-level specific information that could be converted into the coordinates (latitude and longitude).

We identified the four patterns in the trace information that could not be expressed appropriately in PLOD.
First, the releases included a negation of a specific area, such as ``overseas''.
Second, several included cases of the profession information, such as ``a bus guide'' and ``a driver'', indicated the possibility of trips but not did not specify it in a manner sufficient to trace the case.
Figure~\ref{fig:failure} illustrates a case corresponding to this category.
Third, there were the releases that indicated traces in generic terms, for example, ``returning home with a child'', which could not be converted into the coordinates.
Lastly, several releases referred to the other press releases that could contain the trace information, such as ``same as past case A''.
It is hardly possible to convert the first and the second cases into the coordinates, whereas the third and the fourth cases might be processed to obtain the required data.
For example, we may infer that the ``home'' is supposed to be in the city where the press release has been issued, and, similarly, the reference in the fourth case might be appropriately processed.
Both cases can be organized as the best fit for the SPARQL-based queries that can infer the information even if it is not explicitly stated.

Limitations of this evaluation are associated with the sampling bias related to the fact that the analysis has been conducted only on the limited number of samples without the possibility to apply random sampling.
Additionally, the coding has been implemented by a single person, and the statistical analysis is not possible.
Nevertheless, the collected samples can be used to illustrate that the proposed framework is capable of expressing the patient location data described in the current press releases, although there are possibilities for the further improvement.

\begin{figure}
\vspace*{6mm}
\begin{lstlisting}
<https://plod.info/data/29202001281-P01-M01> a schema:MoveAction .
    schema:agent <https://plod.info/data/29202001281-P01> ;
    schema:startTime "2020-01-08"^^schema:DateTime ;
    schema:endTime "2020-01-11"^^schema:DateTime ;
    schema:fromLocation "" ;
    schema:toLocation "" ;
    schema:instrument "Bus" .
\end{lstlisting}
\vspace*{3mm}
  \caption{Example of a failure case}
  \label{fig:failure}
\end{figure}

\section{Discussion}
\label{sec:discussion}

Concerning the actual deployment of the proposed public health service in Japan, we need to execute several steps as follows.
In the Japanese health care system, each local government is responsible for control of infectious diseases, and therefore, public releases are issued by them either in HTML, or PDF format at their websites.
In this setting, there are mainly two ways to provide mobile applications with access to the open data.
The one approach is to establish a centralized model, in which a national public health authority, such as Ministry of Health, would perform the entire processing, such as collection of releases, conversion of formats, and distribution.
The other approach is to implement a decentralized approach in which each local government would issue its announcement in the RDF format using the standardized vocabularies.
In the latter case, there would be a need for a public RDF data repository providing mobile applications with the data they need.
As both approaches necessitate the substantial coordination between organizations, a reasonable option would be to collect, convert, and distribute the open data via a third party organization, considering that the releases are publicly usable.

An indispensable part of the proposed approach is to develop the mobile applications that would use the open data to raise customized alerts for users.
Mobile phones typically have a built-in application programming interfaces (API) for location services, which responds the position data of the mobile phone.
Accordingly, mobile applications may periodically call the API to record their positions, which can be later used to estimate the intersections with the patients mentioned in the open data. Once the application finds that a patient enters in a predefined range corresponding to the phone, it may alert the owner of the phone.

At the moment, the proposed framework is at a draft stage, and there could be steps required to implement before the public use becomes possible.
As mentioned above, public releases have a variety of formats, and they often contain the irregular data, which are hardly possible to express in the current version of PLOD.
To address the problem, continuous improvement for the specification is necessary, however, the formats of press releases need to be also standardized and updated.
To facilitate the process of creating a unified data representation, it would be preferable to launch the service, even with the limited quantity in terms of releases and the limited quality in terms of expressive power.
Then, during the actual use, the specification and application would evolve to meet the real needs.

The proposed application can be implemented in several ways.
It could be developed as an independent application or as an embedded module in an application that uses a location service, such as applications for crisis management provided by local governments.
As the proposed framework is based on the RDF, users or third parties can easily add the information relevant according to their needs.
Organizing a hackathon is an approach to develop a variety of applications suitable for various usage options.
As the public health information concerning infectious diseases may attract attention of a considerable number of people, the proposed application can be widely installed, once the application becomes available for the public use.

Under the pandemic conditions, there could be a surge of data requests, once the applications are widely installed in the population.
Accordingly, the service needs to be scalable aiming to assure the stability of the service as a crisis management system.
In this regard, the open data might be distributed via Content Delivery Networks (CDNs).
Even a state-of-art RDF store may not be suitable to handle the surge of SPARQL queries from a large number of mobile phones.
To mitigate this problem, we may also define a simplified query, equivalent to a full-fledged SPARQL query.
SPARQL provides a flexible framework to retrieve various data at the cost of the overhead to process the complex query on the server side.
Accordingly, instead of issuing a SPARQL query, each application may issue a simplified queries consisting of a couple of arguments so as to accommodate a higher number of requests, while providing the flexibility of SPARQL.

\section{Conclusion}
\label{sec:conclusion}

In this study, we have proposed a framework to represent information press releases focused on the current situation on infected patients in a form of the open data using RDF.
To the best of our knowledge, this is the first attempt of applying the RDF and open data technology to mitigation of epidemic risks.
Due to the ongoing pandemic caused by COVID-19, we consider that the proposed framework has a potential to be used worldwide.

The future research work will include the following three steps.
First, we plan to collect the public releases, convert them into the open RDF data format, and distribute them aiming to facilitate the development of applications.
This could serve as a demonstration for stakeholders to evaluate the utility of the open public health data concept.
Second, the proposed framework needs to be supported by the authorities, such as the ministry of health or a national institution, in a centralized manner.
Third, local governments are expected to initiate providing the data in the linked open data format directly in a decentralized manner.

In the field of epidemiology, public health fields have been using information technologies mostly for the purposes of statistics and prediction; however, they have rarely considered using such methods for prevention of diseases.
Moreover, the concept of open data in the medical field has not been considered widely, and there exist only a limited number of publication dedicated to this subject.
We envisage that the proposed approach may serve as a foundation for development of knowledge processing applications in public health, such as alerting about hospital overload under pandemic conditions.

\section*{Acknowledgment}

The authors thank Ms. Rie Nariko and Ms. Nobuko Nakagomi for their
contribution in the collection of press releases.

\bibliographystyle{IEEEtran}
\bibliography{plod-arxiv}

\begin{thebibliography}{10}
\providecommand{\url}[1]{#1}
\csname url@samestyle\endcsname
\providecommand{\newblock}{\relax}
\providecommand{\bibinfo}[2]{#2}
\providecommand{\BIBentrySTDinterwordspacing}{\spaceskip=0pt\relax}
\providecommand{\BIBentryALTinterwordstretchfactor}{4}
\providecommand{\BIBentryALTinterwordspacing}{\spaceskip=\fontdimen2\font plus
\BIBentryALTinterwordstretchfactor\fontdimen3\font minus
  \fontdimen4\font\relax}
\providecommand{\BIBforeignlanguage}[2]{{%
\expandafter\ifx\csname l@#1\endcsname\relax
\typeout{** WARNING: IEEEtran.bst: No hyphenation pattern has been}%
\typeout{** loaded for the language `#1'. Using the pattern for}%
\typeout{** the default language instead.}%
\else
\language=\csname l@#1\endcsname
\fi
#2}}
\providecommand{\BIBdecl}{\relax}
\BIBdecl

\bibitem{rasmussen2018cdc}
S.~A. Rasmussen and R.~A. Goodman, \emph{The CDC Field Epidemiology
  Manual}.\hskip 1em plus 0.5em minus 0.4em\relax Oxford University Press,
  2018.

\bibitem{yang2009use}
C.~Yang, J.~Yang, X.~Luo, and P.~Gong, ``Use of mobile phones in an emergency
  reporting system for infectious disease surveillance after the sichuan
  earthquake in china,'' \emph{Bulletin of the World Health Organization},
  vol.~87, pp. 619--623, 2009.

\bibitem{kim2017middle}
K.~Kim, T.~Tandi, J.~W. Choi, J.~Moon, and M.~Kim, ``Middle east respiratory
  syndrome coronavirus (mers-cov) outbreak in south korea, 2015: epidemiology,
  characteristics and public health implications,'' \emph{Journal of Hospital
  Infection}, vol.~95, no.~2, pp. 207--213, 2017.

\bibitem{jones2019toward}
K.~H. Jones, H.~Daniels, S.~Heys, and D.~V. Ford, ``Toward an ethically founded
  framework for the use of mobile phone call detail records in health
  research,'' \emph{JMIR mHealth and uHealth}, vol.~7, no.~3, p. e11969, 2019.

\bibitem{ouedraogo2019does}
B.~Ouedraogo, J.~Gaudart, and J.-C. Dufour, ``How does the cellular phone help
  in epidemiological surveillance? a review of the scientific literature,''
  \emph{Informatics for Health and Social Care}, vol.~44, no.~1, pp. 12--30,
  2019.

\bibitem{lai2019measuring}
S.~Lai, A.~Farnham, N.~W. Ruktanonchai, and A.~J. Tatem, ``Measuring mobility,
  disease connectivity and individual risk: a review of using mobile phone data
  and mhealth for travel medicine,'' \emph{Journal of travel medicine},
  vol.~26, no.~3, p. taz019, 2019.

\bibitem{jones2018challenges}
K.~H. Jones, H.~Daniels, S.~Heys, and D.~V. Ford, ``Challenges and potential
  opportunities of mobile phone call detail records in health research,''
  \emph{JMIR mHealth and uHealth}, vol.~6, no.~7, p. e161, 2018.

\bibitem{jahani2017improving}
E.~Jahani, P.~Sunds{\o}y, J.~Bjelland, L.~Bengtsson, Y.-A. de~Montjoye
  \emph{et~al.}, ``Improving official statistics in emerging markets using
  machine learning and mobile phone data,'' \emph{EPJ Data Science}, vol.~6,
  no.~1, p.~3, 2017.

\bibitem{chirombo2018review}
J.~Chirombo, P.~Diggle, D.~Terlouw, and J.~Read, ``A review of models of human
  mobility for predicting infectious disease spread,'' \emph{Modelling spatial
  processes of infectious diseases}, p.~60, 2018.

\bibitem{panigutti2017assessing}
C.~Panigutti, M.~Tizzoni, P.~Bajardi, Z.~Smoreda, and V.~Colizza, ``Assessing
  the use of mobile phone data to describe recurrent mobility patterns in
  spatial epidemic models,'' \emph{Royal Society open science}, vol.~4, no.~5,
  p. 160950, 2017.

\bibitem{sallah2017mathematical}
K.~Sallah, R.~Giorgi, L.~Bengtsson, X.~Lu, E.~Wetter, P.~Adrien, S.~Rebaudet,
  R.~Piarroux, and J.~Gaudart, ``Mathematical models for predicting human
  mobility in the context of infectious disease spread: introducing the
  impedance model,'' \emph{International journal of health geographics},
  vol.~16, no.~1, p.~42, 2017.

\bibitem{wesolowski2016connecting}
A.~Wesolowski, C.~O. Buckee, K.~Eng{\o}-Monsen, and C.~J.~E. Metcalf,
  ``Connecting mobility to infectious diseases: the promise and limits of
  mobile phone data,'' \emph{The Journal of infectious diseases}, vol. 214, no.
  suppl\_4, pp. S414--S420, 2016.

\bibitem{bengtsson2011improved}
L.~Bengtsson, X.~Lu, A.~Thorson, R.~Garfield, and J.~Von~Schreeb, ``Improved
  response to disasters and outbreaks by tracking population movements with
  mobile phone network data: a post-earthquake geospatial study in haiti,''
  \emph{PLoS medicine}, vol.~8, no.~8, 2011.

\bibitem{finger2016mobile}
F.~Finger, T.~Genolet, L.~Mari, G.~C. de~Magny, N.~M. Manga, A.~Rinaldo, and
  E.~Bertuzzo, ``Mobile phone data highlights the role of mass gatherings in
  the spreading of cholera outbreaks,'' \emph{Proceedings of the National
  Academy of Sciences}, vol. 113, no.~23, pp. 6421--6426, 2016.

\bibitem{cinnamon2016evidence}
J.~Cinnamon, S.~K. Jones, and W.~N. Adger, ``Evidence and future potential of
  mobile phone data for disease disaster management,'' \emph{Geoforum},
  vol.~75, pp. 253--264, 2016.

\bibitem{bengtsson2015using}
L.~Bengtsson, J.~Gaudart, X.~Lu, S.~Moore, E.~Wetter, K.~Sallah, S.~Rebaudet,
  and R.~Piarroux, ``Using mobile phone data to predict the spatial spread of
  cholera,'' \emph{Scientific reports}, vol.~5, p. 8923, 2015.

\bibitem{gagnon2016m}
M.-P. Gagnon, P.~Ngangue, J.~Payne-Gagnon, and M.~Desmartis, ``m-health
  adoption by healthcare professionals: a systematic review,'' \emph{Journal of
  the American Medical Informatics Association}, vol.~23, no.~1, pp. 212--220,
  2016.

\bibitem{WSJfeb52020}
L.~Lin, ``{China Marshals Its Surveillance Powers Against Coronavirus},''
  \\https://www.wsj.com/articles/china-marshals-the-power-of-its-surveillance-state-in-fight-against-coronavirus-11580831633.

\bibitem{WSJfeb172020}
E.-Y. Jeong, ``{South Korea Tracks Virus Patients’ Travels—and Publishes
  Them Online},''
  \url{https://jp.wsj.com/articles/SB12291155354026644516304586207690851702666}.

\bibitem{okumura2019ghtc}
T.~Okumura, ``Tracing infectious agents with mobile location information: A
  simple and effective countermeasure against epidemic risks,'' in \emph{2019
  {IEEE} Global Humanitarian Technology Conference ({GHTC} 2019)}.\hskip 1em
  plus 0.5em minus 0.4em\relax IEEE, 2019.

\bibitem{torniai2006sharing}
\BIBentryALTinterwordspacing
C.~Torniai, S.~Battle, and S.~Cayzer, ``Sharing, discovering and browsing photo
  collections through rdf geo-metadata,'' in \emph{{SWAP} 2006 - Semantic Web
  Applications and Perspectives, Proceedings of the 3rd Italian Semantic Web
  Workshop}, ser. {CEUR} Workshop Proceedings, vol. 201.\hskip 1em plus 0.5em
  minus 0.4em\relax CEUR-WS.org, 2006. [Online]. Available:
  \url{http://ceur-ws.org/Vol-201/07.pdf}
\BIBentrySTDinterwordspacing

\bibitem{battle2011geosparql}
R.~Battle and D.~Kolas, ``Geosparql: enabling a geospatial semantic web,''
  \emph{Semantic Web Journal}, vol.~3, no.~4, pp. 355--370, 2011.

\bibitem{beckett2004rdf}
D.~Beckett and B.~McBride, ``Rdf/xml syntax specification (revised),''
  \emph{W3C recommendation}, vol.~10, no. 2.3, 2004.

\bibitem{prudhommeaux2008sparql}
\BIBentryALTinterwordspacing
E.~Prud'hommeaux and A.~Seaborne, ``{SPARQL Query Language for RDF},'' {W3C
  Recommendation}, 2008. [Online]. Available:
  \url{http://www.w3.org/TR/rdf-sparql-query/}
\BIBentrySTDinterwordspacing

\end{thebibliography}

\end{document}